\documentclass{acmsiggraph}
\usepackage{mathptmx}
\usepackage{graphicx}

\title{Embedded Reflection Mapping} 

\author{Paul Anderson\thanks{e-mail: p.anderson@gsa.ac.uk} \\ Digital Design
Studio - Glasgow School of Art \and Gon\c{c}alo
Carvalho\thanks{e-mail: g.carvalho@gsa.ac.uk} \\ Digital Design
Studio - Glasgow School of Art}

\begin{document}

\maketitle

\section{Introduction}

Environment maps are used to simulate reflections off curved objects. We present
a technique to reflect a user, or a group of users, in a real environment, onto a virtual object, in a virtual
reality application, using the live video feeds from a set of cameras, in
real-time.
Our setup can be used in a variety of environments ranging from outdoor or
indoor scenes. 

The need for such a technique becomes obvious under the definition
of {\it replacement reality}: the ability to interact with highly credible digital models in a natural and intuitive manner, unencumbered by intrusive equipment, using your senses of touch and hearing, as well as vision, offering the user an enhanced and personalised immersive experience. Consider a virtual car showroom.  Confronting a user with a virtual customization toolset for a car, providing a high-fidelity virtual representation, enhanced by a sense
of immersion including both surrounding environment and the user within the same scene, is an invaluable marketing tool.

\section{Embedded Reflection Mapping}

Environment maps are textures that describe, for all directions, the incoming or
outgoing light at one point in space. The implicit assumption, with environment
maps, that the incoming illumination on a point in a surface really depends only
on the direction of the reflected ray is still valid \cite{olano:2002:RTS}.

The mirror reflection term, described by an environment map, can be used in combination with local illumination terms. By combining high-dynamic range images and a filtered version of the environment map, lighting effects as described by \cite{debevec:1998:RSO} are possible.

The specular term of a lighting model can be factored in a standard environment map and a fresnel term. Our environment map is achieved by compositing of a real scene environment map and an environment map with the user, or group of users.

Under this factorization, we wish to further factor our environment map. Our standard 
environment map will be a composition of an environment map of a real scene and an environment
map containing only the user or group of users, lit as they would stand in the real scene environment,
and generated using the same camera parameters of our real scene environment, in real-time.

Environment maps represent directionally dependent information. The mapping between directions,
as encoded in the map, and texture coordinates is the parameterization of the environment map.

The parameterization we have chosen is the spherical environment map. The light information on a spherical
environment map for a certain point is all the visible light from that given point in space projected
onto the inside surface located concentrically surrounding that point. The light information can then be
encoded onto a two-dimensional angular plane using the discretized form of the parametric equation of
the sphere.

In order to build such an environment map, the desired capture point, must be captured
along different directions. These images are then warped onto a sphere along the corresponding
directions. The warping is done by sampling the sphere along all of the directions and classifying the 
pixels according to each of the directions.

\begin{figure}[ht]
\centering
\includegraphics[width=2in]{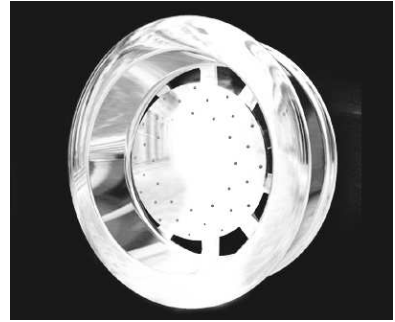}
\caption{User being reflected onto a virtual wheel. The black spots are part of the geometry of the wheel.
The lightprobe on the wheel is courtesy of Paul Debevec.}
\end{figure}

Our camera setup follows closely that of \cite{debevec:2002:ALR}. Color and infrared images of the user
or group of users are captured and the live video feeds are then streamed over to a PC server. 
After the mattes are acquired, the composite environment map is generated and modulated with a Fresnel
term to achieve the final reflection onto the virtual object.
 
\section{Conclusion}

We have described a technique to reflect a user or group of users, dynamically,
onto a virtual object. An example of a virtual
reality application in which the technique is useful has been presented.

It is the authors intent to combine this technique with environment map prefiltering in order to improve the visual quality
of the resulting image.

\bibliographystyle{acmsiggraph}
\bibliography{erm}

\end{document}